# Nanothermodynamics – A generic approach to material properties at nanoscale

**A. K. Rajagopal (1) , C. S. Pande (1) , and Sumiyoshi Abe (2)**
(1) Naval Research Laboratory, Washington D.C., 20375, USA
(2) Institute of Physics, University of Tsukuba, Ibaraki 305-8571, Japan

## ABSTRACT

Granular and nanoscale materials containing a relatively small number of constituents have been studied to discover how their properties differ from their macroscopic counterparts. These studies are designed to test how far the known macroscopic approaches such as thermodynamics may be applicable in these cases. A brief review of the recent literature on these topics is given as a motivation to introduce a generic approach called "Nanothermodynamics". An important feature that must be incorporated into the theory is the non-additive property because of the importance of 'surface' contributions to the Physics of these systems. This is achieved by incorporating fluctuations into the theory right from the start. There are currently two approaches to incorporate this property: Hill (and further elaborated more recently with Chamberlin) initiated an approach by modifying the thermodynamic relations by taking into account the surface effects; the other generalizes Boltzmann-Gibbs statistical mechanics by relaxing the additivity properties of thermodynamic quantities to include nonextensive features of such systems. An outline of this generalization of the macroscopic thermodynamics to nano-systems will be given here.

# I. INTRODUCTION

The question " Small systems: when does thermodynamics apply?" was raised seven decades ago in the context of nuclear reactions [1] and continues to be asked to this day! Because of the advent of sophisticated techniques of creating matter in any desired form, especially in " small" sizes ranging from grains (microns), nanosystems, molecular magnets, and atomic clusters displaying a variety of interesting physical properties, this question has to be asked again. It was clear right from the beginning that as the system size decreases one has to deal with fluctuations. In the context of nuclear reactions, the first such considerations were on temperature fluctuations [1]. Only in 1992, quantitative measurements of temperature fluctuations in a physical system using superconducting magnetometers was reported [2]. With more recent advances in thermometry etc, further investigation may be forthcoming in the near future. An early discussion of "Thermodynamics of small systems" was developed by Hill [3] four decades ago because of his interest in polymers and macromolecules besides his deep inquiry into thermodynamics per se. The most important point of this inquiry is to understand the overall thermodynamic equilibrium behavior of systems in a generic fashion, independent of the detailed microscopic dynamics. It is remarkable that Hill expressed the belief in the Introduction to Part I of his book "Thermodynamics of small systems" in the following way: "The applicability of statistical mechanical ensemble theory to small systems as well as large suggests strongly that a parallel thermodynamics should exist." The one and only example of such a description known at the time of his writing was the theory of thermodynamic equilibrium due to Boltzmann and Gibbs. Hill therefore made a suitable modification of this theory to develop his "Nanothermodynamics". In the last few years, he has returned to this topic in collaboration with Chamberlin [4] due to the current surge of activity in nanometric materials. Recently attention has turned to the non-equilibrium thermodynamics and fluctuations as well due to the works of Gallovotti and Cohen, and that of Jarzynski [5].

A representative set of experimental investigations into physical properties of smaller and smaller system sizes in the recent past may give the reader a flavor of this activity. Accurate determination of surface energy of Ag nanoparticles was found to be significantly higher than the bulk values [6]. The size dependence of surface ferromagnetism of Pd fine particles was found only on (100) facets [7]. Detection of the $\sqrt{N}$ statistical polarization in a small ensemble of electron spin centers was by magnetic resonance force spectroscopy [8]. An unexpected decrease in the strength of various materials as sizes of microns (grains) become nanometers [9] was found. A large critical current at nanosizes of superconductors [10] was measured. There are experiments on molecular magnets [11] and possible canted magnetism in Gd clusters [12]. There are also discussions of microelasticity at the nanoscale [13]. There are also recent theoretical works dealing with existence of temperature on the nanoscale [14], possible tests of the thermodynamic approach to granular media [15], and scaling relations in equilibrium thermostatistics with long range interactions with implications to nanosystems [16].

The brief description of this topic given here suffices to provoke interest in this controversial topic. In this paper, we wish to present a purely " macro" view of the subject matter without directly invoking the microscopic underpinning of the phenomena arising out of the reduced size of the system as far as possible. This is in the spirit of the thermodynamic approach to a general understanding of equilibrium and near equilibrium properties of matter. When one reaches the system size where one has a collection of very small number of particles where the details of interactions among particles are important for the description of the system, a quantum statistical mechanical approach becomes essential. A thermodynamic – like description may then be deduced therefrom with suitable introduction of quantities pertaining to the system properties. Our presentation of the "nanothermodynamics" will be based on a different formulation of the
" ensemble of small systems approach" that Hill originally proposed. The inherent fluctuations in the system are incorporated which naturally leads to a modified Boltzmann-Gibbs scheme of the form suggested by Tsallis [17]. An important consequence of this is that the entropy of the system is not additive which was the case with the Hill procedure.

In the next section, we give a brief review of the traditional thermodynamics to set the stage for presenting the Hill theory of nanothermodynamics in sec.III. In the following sec.IV, our theory is given. In the final sec.V, concluding remarks are made.

## II. A BRIEF REVIEW OF THERMODYNAMICS

Let us recapitulate some basic features of equilibrium thermodynamics by way of introducing the ensemble theory of small systems as was originally done by Hill. The possibility of inclusion of fluctuations will become evident. Consider first a small system with internal energy e and entropy s, immersed in a large one, called the reservoir, with internal energy, E and entropy, S, and the two together are assumed to be isolated. Then since the internal energies and the entropies are additive, any deviations of these quantities are so as to preserve these features and thus the small system attains the temperature, $T = \delta S/\delta E$ of the reservoir ($\delta s/\delta e = \delta S/\delta E$). Similarly, if the volumes v, V of the two are considered, we obtain the equalities $\delta s/\delta v = \delta S/\delta V \equiv P/T$, where P is the pressure. If the number of particles n, N are considered, we obtain $\delta s/\delta n = \delta S/\delta N = -\mu/T$, where $\mu$ is called the chemical potential. If we consider the number of possible microstates $\Omega(E)$ of the reservoir, the Gibbsian way of understanding the same situation posed above is to identify the entropy of the reservoir to be $S = k \ln \Omega(E)$, where k is the Boltzmann constant. Then the probability of finding the small system with energy e is given by $f(e) = \Omega(E-e)/\Omega(E) \approx \exp - e/kT$, after using the fact that e is much smaller than E. This is the well-known Boltzmann-Gibbs distribution associated with the small system.

In order to bring out the salient features of equilibrium thermodynamics in relation to the nano-thermodynamics of Hill, we continue further with a brief description of the foundations of the thermodynamics of macroscopic systems. This will enable us to present the generalization to nano-systems in a transparent way. The equilibrium thermodynamic behavior of a macroscopic system is described by an extensive state function entropy, S, which for a one-component system is a function of the extensive variables, the internal energy, U, volume, V, and the number of particles, N.  The extensive property of a macroscopic system is expressed by the statement that any one of these variables is a homogeneous function of degree one of the other three variables. A natural development of the thermodynamic description of this system traditionally starts by considering the internal energy function, $U(S, V, N)$ and its first derivatives defining the intensive conjugate variables, the chemical potential, $\mu = (\partial U/\partial N)_{S,V}$, its temperature, $T = (\partial U/\partial S)_{N,V}$, and pressure, $P = -(\partial U/\partial V)_{S,N}$. The extensive property of $U(S, V, N)$ as a homogeneous function of degree one of the three variables, S, V, N implies that we have an explicit representation

$$U(S, V, N) = TS - PV + \mu N. \tag{1}$$

The first law of thermodynamics for a quasi-static process is often contained in the differential (Pfaffian) form

$$đQ = TdS = dU + PdV - \mu dN. \tag{2}$$

By considering the differential form of eq.(1) and using eq.(2), (or equivalently, taking the external derivative of eq.(1) to be zero in conjunction with the internal) we obtain the celebrated Gibbs – Duhem relation

$$-SdT + VdP - Nd\mu = 0. \tag{3}$$

This relation is important in exhibiting the fact that the three intensive triplet of variables $(\mu, T, P)$ are not independent and the usual choice (T, P) is made in the literature, defining an equation of state for the system. Besides U, three other functions were found to be very useful in the applications to specific physical situations, enthalpy, H(S, P, N) = U(S, V, N)+PV, Helmholtz free energy, F(T, V, N) = U(S, V, N) - TS, and Gibbs free energy, G(T, P, N) = U(S, V, N) -TS+PV. These are all equivalent in describing the system under different physical situations and thus are useful in different experimental situations. The inter-dependence of these functions and their continuity properties in their appropriate variables yield the four thermodynamic relations called the Maxwell relations. In the nanothermodynamics, this is modified because the nanosystem is sensitive to the environment it is placed in, as will be described presently.

### III. NANOTHERMODYNAMICS –HILL'S THEORY

The Hill procedure is an inquiry into the different thermodynamic properties of nano-systems, starting with only the first law of thermodynamics given by eq.(2) and no other feature. This is because the first law of thermodynamics is context independent, which is just another facet of the principle of conservation of energy, based solely on physical considerations of heat and work involved in any quasi – static process. This is directly related to three independent variables U, V, N for a one-component nano-system (henceforth we consider only this in the interest of simplicity of presentation) in contrast to two in the macroscopic system, because we no longer have the extensive property.

In a one-component nano-system considered here the internal energy is not extensive in the variable N and hence the chemical potential will depend on the number of particles, N, in it. Hill [3] expressed this by restating eq.(1) to reflect this feature by introducing a new function, W(T,P,μ), called " subdivision energy" defined by

$$W = U - TS + PV - \mu N. \tag{4}$$

The differential form of this relation, after taking into account eq.(2), the first law of thermodynamics in differential form, leads to the result

$$dW = -SdT - Nd\mu + VdP. \tag{5}$$

In the macroscopic context, eq.(4) would be identically zero, while eq.(5) is the Gibbs – Duhem relationship. These are the first steps in the proposed generalization of thermodynamics to nano-systems. The next important point made by Hill is that the nanosystem is sensitive to environment. For example, the nanosystem of N items in a volume V immersed in a heat bath at temperature T is different from the same system in contact with a reservoir with chemical potential μ chosen such that the mean number $\overline{N}$ is the same numerical value as N. In the nanosystem, the fluctuations are important unlike in the macrosystems. To take account of these situations, Hill introduces the subdivision energy, W, as above. The rest of the development follows the traditional path.

### IV. NANOTHERMODYNAMICS – OUR THEORY

Our presentation differs from Hill's approach by recognizing that when we put the ensemble of nanosystems in contact with the reservoir, each of the nanosystem *fluctuates* around the temperature of the reservoir, for example. As pointed out by Hill, we could also consider other " environmental" situations where the corresponding entity (e. g. pressure or chemical potential) fluctuates. The origin of these fluctuations lies in the very same nanosize and thus they come to quasi-thermodynamic equilibrium with the reservoir. Such an idea is already present in the description of thermodynamics of small systems by Feshbach[1] in the context of nuclear reactions. This means that the Boltzmann-Gibbs distribution mentioned at the beginning of this description has to be averaged over the temperature fluctuations induced by the reservoir, in this example. Only within the last four years this idea has been further developed using a noisy reservoir [18, 19] in different physical contexts and more generally, using the mathematical theory of large deviations statistics [20]. To make this point more transparent, let us consider Hill's ensemble of nanosystems, whose mean total internal energy is

$$U_n = \frac{1}{n} \sum_{i=1}^{n} u_i . \tag{6}$$

Since these are coupled to the reservoir, each of the $u_i$'s are random variables, implying that $U_n$ is also a random variable. It is assumed that every one of the nanosystems is coupled to the same reservoir, the joint probability distribution is the Gibbsian,

$$P_n = \exp{-\beta_0 n U_n} / Z_n(\beta_0) \tag{7}$$

where $\beta_0$ is the inverse temperature of the bath as in the discussion given earlier. Since the ensemble contains independent nanosystems, each of which is in thermal contact with the reservoir, it also follows from eq.(6) that

$$P_n = p(u_1) p(u_2) \cdots p(u_n) = \frac{e^{-\beta_0 u_1}}{Z(\beta_0)} \frac{e^{-\beta_0 u_2}}{Z(\beta_0)} \cdots \frac{e^{-\beta_0 u_n}}{Z(\beta_0)} \qquad (8)$$

It is here that we depart from Hill and consider possible fluctuations in temperature in determining the joint probability. It is here the mathematical theory of large deviations approach [20] comes in handy in deducing the joint probability. More physically, the result may be expressed as merely taking an integral over all possible fluctuating inverse temperatures that are $\chi^2$-distributed:

$$\int_0^\infty d\beta \, e^{-\beta u} \, f(\beta) = \left(1 + (q-1)\beta_0 u\right)^{-\frac{1}{q-1}} \qquad (q > 1) \qquad (9)$$

with

$$f(\beta) = \frac{1}{\Gamma\left(\frac{1}{q-1}\right)} \left\{\frac{1}{(q-1)\beta_0}\right\}^{\frac{1}{q-1}} \beta^{\frac{1}{q-1}-1} \exp\left[-\frac{\beta}{(q-1)\beta_0}\right] \qquad (10)$$

The chi squared distribution is a universal distribution that occurs in many common circumstances such as if $\beta$ is the sum of squares of n Gaussian random variables, with $n = 2/(q-1)$. The constant $\beta_0$ in eq.(10) is the average of the fluctuating $\beta$, $\beta_0 = \int_0^\infty d\beta \, \beta \, f(\beta)$ and the parameter q is the dispersion of $\beta$ given by

$$q - 1 = \beta_0^{-2} \int_0^\infty d\beta (\beta - \beta_0)^2 f(\beta). \qquad (11)$$

When the fluctuation is zero, we recover the usual Boltzmann-Gibbs distribution with q=1 in the above expressions. The distribution function in eq.(9) is the Tsallis distribution! A point of interest that emerges from this is that the associated entropy is the non-additive Tsallis entropy [17], given by $S_q = \left(1 - \sum_i p_i^q\right)/(q-1)$, which when q=1 goes over to the usual additive Gibbs entropy. We may remark that a dynamical reasoning behind the fluctuation may be thought of as arising from some kind of Brownian dispersion caused by the interaction of the heat bath on the nanosystem [18, 19]. There is a thermodynamics that goes with Tsallis entropy [17] and thus we have here an alternate way of describing the nanothermodynamics.

The several environmental effects that Hill considers regarding the nanosystem can be encompassed in a parallel fluctuation theory in the same fashion as for temperature. The corresponding q – values will be associated with the fluctuating entity that is under consideration.

## V. CONCLUDING REMARKS

In conclusion, we have here developed a formal thermodynamic theory of nanosystems incorporating the fluctuations inherent in a nanosystem. This theory naturally leads to a non-Gibbsian theory in contrast to that of Hill, who directly modified the thermodynamic relations due to environmental effects. The nonextensive approach has a modified thermodynamic relations [17]. The development of the nanothermodynamic theories provide both tools to guide experiments and a basis for testing the foundations.

**ACKNOWLEDGEMENTS**
AKR thanks Professor Chamberlin for drawing attention to his work in collaboration with Professor Hill and providing him with several publications on the subject of nanothermodynamics. We thank Dr. Ron Rendell for reading the draft of the paper and suggesting improvements of the presentation. AKR also thanks the Indo–US Workshop for providing him the full support for participating in the meeting entitled "Nanoscale materials: From Science to Technology" held at Puri, India. The first two authors are supported in part by the Office of Naval Research. SA was supported in part by the Grant-in-Aid for Scientific Research of Japan Society for the Promotion of Science.